# Self-averaging stochastic Kohn-Sham density functional theory

Roi Baer[*], Daniel Neuhauser[†] and Eran Rabani[‡]

[*] *Fritz Haber Center for Molecular Dynamics, Institute of Chemistry, Hebrew University, Jerusalem 91904 Israel.*
[†] *Department of Chemistry and Biochemistry, University of California, Los Angeles CA 90095, USA.*
[‡] *School of Chemistry, The Sackler Faculty of Exact Sciences, Tel Aviv University, Tel Aviv 69978, Israel.*

We formulate the Kohn-Sham density functional theory (KS-DFT) as a statistical theory in which the electron density is determined from an average of correlated stochastic densities in a trace formula. The key idea is that it is sufficient to converge the *total energy per electron* to within a predefined statistical error in order to obtain reliable estimates of the electronic band structure, the forces on nuclei, the density and its moments, etc. The fluctuations in the total energy per electron are guaranteed to decay to zero as the system size increases. This facilitates "self-averaging" which leads to the first ever report of *sublinear* scaling KS-DFT electronic structure. The approach sidesteps calculation of the density matrix and thus is insensitive to its evasive sparseness, as demonstrated here for silicon nanocrystals. The formalism is not only appealing in terms of its promise to far push the limits of application of KS-DFT, but also represents a cognitive change in the way we think of electronic structure calculations as this stochastic theory seamlessly converges to the thermodynamic limit.

Density functional theory is used for studying properties of condensed phase, biological and molecular systems. While the actual goal of DFT is to determine the electronic density and ground state energy, the common approach goes through the Kohn-Sham (KS) equations i.e. calculating, and storing *all* the occupied KS orbitals or the idempotent density matrix. Linear scaling methods were developed to circumvent the need for KS orbitals [1-16], but their reliance on Kohn's "nearsightedness" principle makes them sensitive to the dimensionality and the character of the system.

A complimentary DFT formulation is based on orbital-free methods [17-19], which promise linear-scaling in algorithmic complexity regardless of the dimensionality of the system. However, the imperfect representation of the kinetic energy in these approaches may hamper their accuracy. A recent promising development along these lines, based on the availability of an approximate functional connecting the ground state density of non-interacting electrons to the underlying potential, provides an estimate of the kinetic energy from adiabatic connection integration [20].

In this letter we develop a theory that allows an in-principle accurate bridge between the Kohn-Sham and the orbital-free approaches to DFT, gaining from both worlds. The core idea is to use a stochastic technique to DFT (SDFT) calculating directly the density from the KS Hamiltonian using a trace formula without computing KS orbitals or density matrices. The use of a stochastic technique combined with DFT has been put forward several decades ago based on path integral Monte Carlo [21-24], but has not found general use due to algorithmic problems [21]. In contrast, our approach uses a *deterministic* Chebyshev expansion of the density matrix projection operator and applies it to *stochastic orbitals* thereby circumventing the pathologies of path integral Monte Carlo.

The stochastic aspect guarantees an unbiased statistical error distribution and thus enables us to demand convergence, not in the total energy as typically required, but in the *total energy per electron*. Statistical mechanics guarantees that in the thermodynamic limit the fluctuations in the total energy per electron are negligible. This expected decrease of fluctuations is the basis for the new concept of "self-averaging" introduced in the present electronic structure theory. It also enables the development of an algorithm with *sub-linear* scaling in computational complexity. Thus, the present approach provides for a smooth transition from traditional quantum methods originally developed for small systems to the realm of macroscopic structures understood in terms of ensembles and statistical fluctuations.

Related approaches have been recently developed by us for estimating the rate of multiexciton generation in nanocrystals (NCs) [25], for a linear scaling calculation of the exchange energy [26], for overcoming the computational bottleneck in Møller-Plesset second order perturbation theory [27], and for calculating the Random-Phase-Approximation correlation energy for DFT [28]. These are corrections to the fundamental problem of describing the electron density and total energy, which is addressed here for the first time.

In DFT, given a potential $v(\boldsymbol{r})$, the electron density can be calculated as a trace:

$$n_\mu(\boldsymbol{r}) = tr[\theta_\beta(\mu - \hat{H})\hat{n}(\boldsymbol{r})], \qquad (1)$$

where $\theta_\beta(E) = \frac{1}{2}\mathrm{erfc}[-\beta E]$ is a smoothed Heaviside step function with $\beta E_g \gg 1$, $E_g$ is the frontier orbital gap, $\hat{H} = -\frac{1}{2}\nabla^2 + v(\hat{\boldsymbol{r}})$ is the one-body Hamiltonian (atomic units are used throughout), and $\hat{n}(\boldsymbol{r}) = \delta(\boldsymbol{r} - \hat{\boldsymbol{r}})$. The number of electrons is determined by the chemical potential parameter $\mu$ and its value ($\mu_*$) is chosen to yield the desired number of electron in the system ($N_e$):

$$N_e = \int n_{\mu_*}(\boldsymbol{r})d^3r = tr[\theta_\beta(\mu_* - \hat{H})]. \qquad (2)$$

In the usual approach to KS-DFT the trace is performed using the lowest eigenstates $\psi_n(\boldsymbol{r})$ of $\hat{H}$, where $\theta_\beta(\mu - \hat{H})$ projects on the occupied space so the density is given by $n(\boldsymbol{r}) = \sum_{n=1}^{N_e}|\psi_n(\boldsymbol{r})|^2$. The cubic scaling of conventional KS-DFT methods results from the need to find the $N_e$ lowest energy eigenstates of the Hamiltonian.

In order to eliminate the need for determining the KS eigenstates we use stochastic wave-functions, $\chi(r)$, for which the completeness property holds:

$$\langle |\chi\rangle\langle\chi|\rangle_\chi = \hat{I}, \quad (3)$$

where $\hat{I}$ is the identity operator and $\langle\cdots\rangle_\chi$ represents a statistical expectation value (mean) over the stochastic wavefunctions. From equations (1) and (3) the density is given by:

$$n_\mu(r) = \left\langle |\xi_\mu(r)|^2 \right\rangle_\chi \quad (4)$$

where $\xi_\mu(r)$ is a stochastic *occupied* orbital evaluated using a Chebyshev expansion [15, 29-31]:

$$|\xi_\mu\rangle = \theta_\beta(\mu - \hat{H})|\chi\rangle \approx \sum_{n=0}^{N_C} a_n(\mu)|\varphi_n\rangle. \quad (5)$$

In the above, $N_C$ is the length of the Chebyshev polynomial, $|\varphi_n\rangle$ is a fixed linear combination of $|\varphi_{n-1}\rangle$, $\hat{H}|\varphi_{n-1}\rangle$ and $|\varphi_{n-2}\rangle$, $\varphi_0 = \chi$ and $\varphi_{-1} = 0$ (see details in Ref. [32]). The Chebyshev expansion length $N_C$ is proportional to $\beta\Delta E$, where $\Delta E$ is the energy range of the Hamiltonian $\hat{H}$ and is of the order $10^3 - 10^4$. Note that while most of the numerical effort goes into the calculation of $\varphi_n$, these states do not depend on the chemical potential $\mu$. Only the coefficients $a_n(\mu)$ depend on it. Thus it is not significantly more expensive to determine $|\xi_\mu\rangle$ for several values of $\mu$ than it is for just one. Stochastic wave functions have been used before for estimating the density of states [33] but the present work is the first to use them for computing the electronic spatial density with an SCF KS-DFT procedure.

Since $\langle \xi_\mu(r)\xi_\mu(r')^* \rangle_\chi$ is the KS density matrix, any one-particle observable (e.g. the kinetic energy, the local and non-local pseudopotential energy, forces on the nuclei, etc.) is computed as an average:

$$\langle\hat{A}\rangle_\mu = \left\langle \langle\xi_\mu|\hat{A}|\xi_\mu\rangle \right\rangle_\chi \quad (6)$$

In the SDFT calculation for $N_e$ electrons, one starts by guessing an initial density $n(r)$ and the chemical potential $\mu_*$. For example, the initial density can be taken as the sum over the spherical densities of the atoms. The following steps are then used to update the density and the chemical potential:

- Compute the KS potential from the density.
- Generate stochastic orbitals. In a grid (or plane-wave) representation $\chi(r_g) = e^{i\theta_g}/\sqrt{h^3}$ where $\theta_g$ is a random phase and $h$ is the grid spacing. For each orbital, calculate its projection onto the occupied space $\xi_\mu(r)$ using Eq. (5).
- Determine the density $n_\mu(r)$ from Eq. (4) for several values of $\mu$ using the same Chebyshev expansion. Linearly interpolate to find the correct value of the chemical potential, $\mu_*$ using Eq. (2).
- Reiterate until $\mu_*$ converges below a defined threshold.

It is important to note that one must keep the same random orbitals throughout the self-consistent procedure. Otherwise, the convergence of the self-consistent procedure would be limited by the stochastic noise. In addition, while the above procedure is exact in the limit of an infinite set of $\chi's$, in practice, one uses a finite set containing $I$ stochastic orbitals and the estimates of the density and the expectation values involve a statistical error (SE) that is proportional to $\frac{1}{\sqrt{I}}$.

Table 1: Parameters for the silicon nanocrystals. Shown, are the number of grid points in each Cartesian dimension $N_g^{1/3}$, the $\beta$ parameter used to represent the Heaviside function, the length of the Chebyshev expansion ($N_C$), the total energy per electron from the deterministic LDA calculation (when available) and the corresponding stochastic result (based on $I$ iterations).

| System | $N_g^{1/3}$ | $E_h\beta$ | $N_C$ | $I$ | $E/N_e$ (eV) Deterministic | $E/N_e$ (eV) Stochastic |
|---|---|---|---|---|---|---|
| $Si_{16}H_{16}$ | 42 | 55 | 4270 | 2700 | −24.023 | −24.025 |
| $Si_{35}H_{36}$ | 60 | 55 | 4560 | 2700 | −23.921 | −23.919 |
| $Si_{87}H_{76}$ | 64 | 80 | 6330 | 2700 | −23.550 | −23.545 |
| $Si_{353}H_{196}$ | 96 | 125 | 9530 | 720 | --- | −23.845 |
| $Si_{705}H_{300}$ | 108 | 155 | 11510 | 180 | --- | −23.448 |

We demonstrate the approach on a series of silicon nanocrystals (NCs) described in Table 1. The calculations were done using the local density approximation (LDA) for the XC energy [34] and the nuclear electron interaction was described using Troullier-Martins norm conserving pseudopotentials [35] with a local and nonlocal part: $\hat{V} = v_L(r) + \hat{V}_{NL}$. The total energy

$$E = \left\langle \langle\xi|\hat{T} + \hat{V}_{NL}|\xi\rangle \right\rangle_\chi + \int n(r)v_L(r)d^3r \\ + \frac{1}{2}\iint \frac{n(r)n(r')}{|r-r'|}d^3rd^3r' + E_{XC}[n] \quad (7)$$

was converged self-consistently to $10^{-7}E_h$ per electron using DIIS acceleration [36]. The last two columns in Table 1 show several meV deviations of the stochastic energy per particle from the exact LDA value. The expectation value was estimated from 15 samples of independent SCF runs, each with a different set of $I = 1200$ stochastic orbitals. The SE is the square root of the variance of this random variable and is estimated from the standard deviation of the sample.

In Figure 1 we show the SE as a function of system size. As the system size increases the SE decreases. This can be explained by considering a large system composed of two non-interacting identical parts $A + B$ having $N_A$ and $N_B$ electrons. The SE in the total energy of the two independent parts relates to the SE in the total energy of each part as $\sigma_t^2 = \sigma_A^2 + \sigma_B^2 = 2\sigma_A^2$. Thus, for large systems the SE of the total energy scales as $\sigma_t^2 \propto N_e$ where $N_e = N_A + N_B$ is the total number of electrons. For the statistical error of the energy *per particle* one then finds that $\sigma^2 = \left(\frac{\sigma_t}{N_e}\right)^2 \propto \frac{1}{N_e}$, *i.e.* inversely proportional to the system size. Thus the SE per electron is expected to decrease as $N_e^{-1/2}$ for large system. Figure 1 shows that this expected behavior holds for systems with $N_e > 200$.

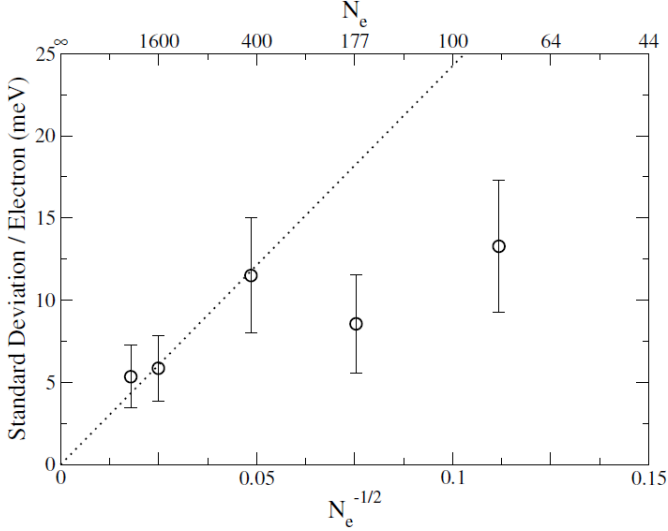

Figure 1: The statistical error per electron (for $I = 1200$ stochastic orbitals) as a function of $N_e^{-1/2}$ where $N_e$ is the number of electrons. The dotted line is a fit to the relation $\sigma = aN_e^{-1/2}$ with $a \approx 250$meV.

Figure 2 shows the CPU time for the SDFT as a function of system size for a fixed SE per electron of 10 meV. We find that the algorithmic scaling is *sub-linear* with system size. The total CPU time depends on several factors: The grid size which scales linear with the number of electrons; the inverse HOMO-LUMO gap which determines the values of $\beta$ and the length of the Chebyshev expansion $N_c$ and roughly scales as $E_b + \Delta/N_e^{2/3}$ ($E_b$ is the bulk band gap and $\Delta$ is a constant); and the number of stochastic orbitals needed to reach a given SE per electron.

The fact that the SE per electron decreases as $1/\sqrt{N_e}$ implies that, for a fixed SE per electron, fewer stochastic orbitals are required to converge the expectation values of the energy per electron to within a desired tolerance as the system grows (see inset of Figure 2). In the limit of a macroscopic system only a single stochastic orbital will be needed to estimate the total energy per electron with negligible error. This "self-averaging" property leads to sub-linear scaling at intermediate system sizes. Once the system size is so large as to enable a single stochastic orbital to be used (for the present system, this is extrapolated to be at 500,000 electrons), the numerical effort of the approach will scale linearly with system size (regime not shown).

In comparison to the deterministic approach, the CPU time for the SDFT calculation crosses that of the deterministic calculation at systems containing about 3000 electrons (order of 700 silicon atoms) for a SE per electron of 10meV. It is interesting to note that the calculations shown in Figure 1 and Figure 2 were carried on a parallel cluster (with 180 cores) for the stochastic approach and on a single core for the deterministic one. As the communication overhead is negligibly small, such a comparison can be made without any bias towards the single-core calculation. This is another significant advantage of the present theory – it scales linearly with the number of cores as long as this number is smaller than $I$. The scaling of the present approach with CPU time and memory is unprecedented for 3D structures. Once again, it relies on converging the total energy per electron rather than the traditional approaches to converge the total energy.

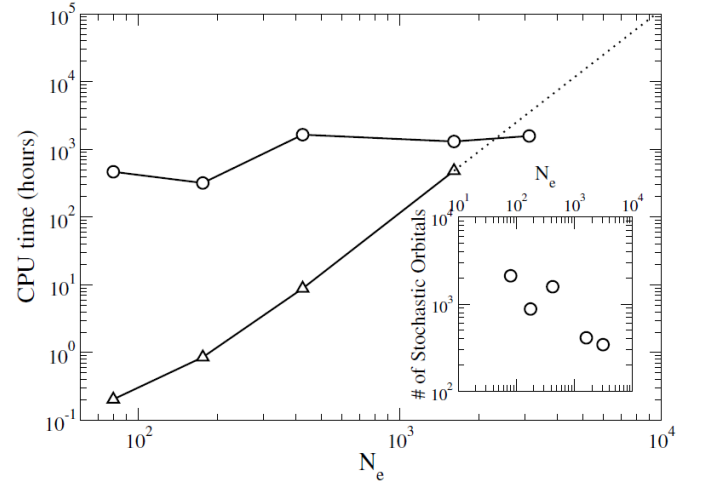

Figure 2: CPU time for full (triangles) and stochastic (circles) SCF DFT calculation for silicon NCs as a function of the number of electrons. Inset: Log-log plot of the number of stochastic orbitals required to converge the total silicon NCs energy per electron to 10meV as a function of the number of electrons.

The convergence of the total energy per electron is perhaps sufficient to obtain an accurate estimate of this property alone. However, the SE in the total energy itself is rather high, far from a desired accuracy of tens of meVs. This raises a concern whether the current approach can be used as a reliable tool to obtain expectation values of quantities that depend on all electrons. To address this concern, in Figure 3 we plot the forces along an arbitrary ($x$) direction obtained by the SDFT for one of the silicon NCs studied. Calculations for the other NCs provide a similar picture. The forces were obtained by integrating the density with the gradient of the electron-nucleolus potential.

The two panels in Figure 3 correspond to two sets of SDFT calculations in which a small ($I = 80$) and a larger ($I = 320$) number of stochastic orbitals were used. The SE in the total energy is ~6eV and ~3eV in the two calculations, respectively, yet the error in the forces is of statistical nature and is perfectly controllable: it drops by a factor 2 when $I$ increases by a factor of 4. The SE in the force on the silicon atoms (low index in the figure) is about is 4 times larger than that on the H atoms (high index in the figure). This is due to the higher number of electrons near the silicon nucleus (also by a factor of 4) and could likely be reduced with importance sampling techniques, as will be explored in future studies. The error bar on the force on a silicon atom when the SE in the total energy per electron is 10meV is ~$0.02 E_h/a_0$ such a value is low enough to enable Langevin dynamics simulations which shed light on the equilibrium and non-equilibrium statistical mechanical properties of large systems.

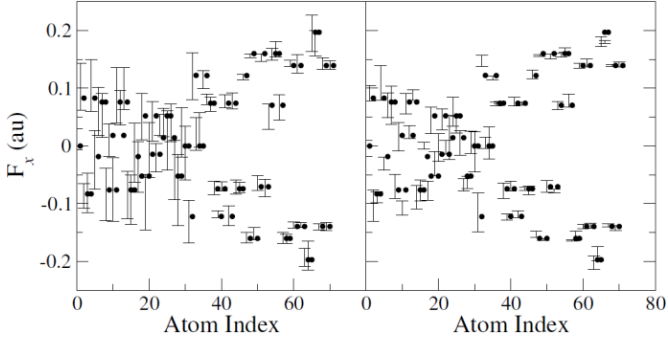

Figure 3: The exact (black full circles) and the 70% confidence interval for the $x$ component of the force on each of the Si (index 0 to 35) and H (index 36 to 70) nuclei in $Si_{35}H_{36}$. The left and right panel use 80 and 320 stochastic orbitals, respectively.

Despite the orbital-less nature of the calculation, the density of states (DOS) and KS orbital energies are readily available. In the left panel of Figure 4 we show the DOS for $Si_{35}H_{36}$ for energies close to the chemical potential $\mu_*$ estimated from a finite difference formula applied to $\partial N_e(\mu)/\partial \mu$. Results obtained from two different stochastic runs are compared to those of the direct approach using the KS orbital gies $N_e^{direct}(\mu_m) \equiv \sum_n \frac{1}{2} \text{erfc}(-\beta(\mu_m - \epsilon_n))$. The value of the chemical potential, $\mu_*$ for each stochastic run varies ($\mu_*$ typically "sticks" to near to the HOMO or to the LUMO of the system) but the DOS is similar between the different runs and agrees well with that of the direct approach, to within a small statistical error.

In the right panel of Figure 4 we compare the value of the KS orbital energies for the same NC. The agreement between the stochastic and deterministic approach is remarkable. The inset of Figure 4 shows the KS orbital energy error defined by the difference between the stochastic and deterministic estimates near the top of the valance band and the bottom of the conduction band. The small error ($< 0.1$eV) has a stochastic and deterministic components. The former can be decreased by increasing the number of stochastic orbitals used for the SDFT procedure while the latter is controlled by the length of the Chebyshev polynomial.

In summary, we have presented a stochastic formulation of DFT that provides a link between KS and orbital-free DFT formulations. The electron density is given in terms of a trace formula which is evaluated using stochastic occupied orbitals generated by a Chebyshev expansion of the occupation operator, rather than by finding all the KS orbitals. Due to the statistical nature of the formalism, the electron density and expectation values involve an unbiased statistical error, which obeys normal large-number statistics and thus, can be controlled by increasing the number of stochastic orbitals $I$. We demonstrated the method and its properties on silicon NCs.

A novel aspect of our formalism is that it allows to control the total energy *per electron* instead of the total energy itself while obtaining statistically significant estimates of electron densities forces on nuclei and band structure from the calculation. For this reason, the method seamlessly connects with the thermodynamic limit where the statistical fluctuations in the total energy per electron vanish. Therefore, our formalism enjoys an effect of "self-averaging", whereas the silicon NC grows one requires fewer stochastic iterations for reaching the same accuracy. It is this effect which enables *sublinear* scaling, never before seen in KS-DFT electronic structure.

Finally, since the method produces a KS Hamiltonian, it can be integrated with the MP2 and RPA stochastic approaches we recently developed [27, 28].

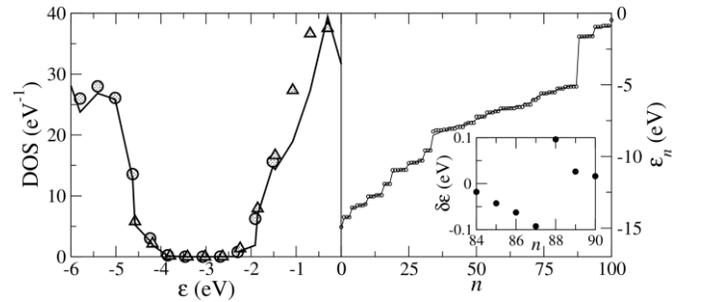

Figure 4: Left: The LDA density of states of $Si_{35}H_{36}$ (black line) and the stochastic estimate for 2 different runs (circles and triangles). $I = 720$ stochastic orbitals. Right: The KS orbital energy, $\varepsilon_n$, versus the index of the orbital, $n$ for the deterministic (black solid) and stochastic estimate (circles). Inset: The KS orbital energy error near the top of the valance (index 87) and bottom of the conduction (index 88) bands.

RB and ER gratefully thank the Israel Science Foundation, grant numbers 1020/10 and 611/11, respectively. DN's work was supported as part of the Molecularly Engineered Energy Materials (MEEM), and Energy Frontier Research Center funded by the DOE, Office of Science, Office of Basic Energy Sciences under Award Number DE-SC0001342.

[1] Y. Wang, G. M. Stocks, W. A. Shelton, D. M. C. Nicholson, Z. Szotek, and W. M. Temmerman, Phys. Rev. Lett. **75**, 2867 (1995).
[2] E. Schwegler and M. Challacombe, J. Chem. Phys. **105**, 2726 (1996).
[3] D. R. Bowler, M. Aoki, C. M. Goringe, A. P. Horsfield, and D. G. Pettifor, Modelling and Simulation in Materials Science and Engineering **5**, 199 (1997).
[4] S. Goedecker, Rev. Mod. Phys. **71**, 1085 (1999).
[5] G. E. Scuseria, J. Phys. Chem. A **103**, 4782 (1999).
[6] J. M. Soler, E. Artacho, J. D. Gale, A. Garcia, J. Junquera, P. Ordejon, and D. Sanchez-Portal, J. Phys.: Condens. Matter. **14**, 2745 (2002).
[7] C. K. Skylaris, P. D. Haynes, A. A. Mostofi, and M. C. Payne, J.

Phys.: Condens. Mater. **17**, 5757 (2005).
[8] M. J. Gillan, D. R. Bowler, A. S. Torralba, and T. Miyazaki, Comp. Phys. Comm. **177**, 14 (2007).
[9] T. Ozaki, Phys. Rev. B **82**, 075131 (2010).
[10] A. Thiess, R. Zeller, M. Bolten, P. H. Dederichs, and S. Blügel, Phys. Rev. B **85**, 235103 (2012).
[11] Z. Rudolf, J. Phys.: Condens. Mater. **20**, 294215 (2008).
[12] E. Rudberg, E. H. Rubensson, and P. Sałek, J. Chem. Theor. Comp **7**, 340 (2010).
[13] C. Ochsenfeld, J. Kussmann, and F. Koziol, Angew. Chem. Int. Edit. **43**, 4485 (2004).
[14] A. H. R. Palser and D. E. Manolopoulos, Phys. Rev. B **58**, 12704 (1998).
[15] R. Baer and M. Head-Gordon, Phys. Rev. Lett. **79**, 3962 (1997).
[16] W. T. Yang, Phys. Rev. Lett. **66**, 1438 (1991).
[17] Y. A. Wang, N. Govind, and E. A. Carter, Phys. Rev. B **60**, 16350 (1999).
[18] T. A. Wesolowski and A. Warshel, J. Phys. Chem. **97**, 8050 (1993).
[19] A. Coomar, C. Arntsen, K. A. Lopata, S. Pistinner, and D. Neuhauser, J. Chem. Phys. **135**, 084121 (2011).
[20] A. Cangi, D. Lee, P. Elliott, K. Burke, and E. K. U. Gross, Phys. Rev. Lett. **106**, 236404 (2011).
[21] G. G. Hoffman, L. R. Pratt, and R. A. Harris, Chem. Phys. Lett. **148**, 313 (1988).
[22] R. A. Harris and L. R. Pratt, J. Chem. Phys. **82**, 856 (1985).
[23] W. T. Yang, Phys. Rev. Lett. **59**, 1569 (1987).
[24] W. T. Yang, Phys. Rev. A **38**, 5504 (1988).
[25] R. Baer and E. Rabani, Nano Lett. **12**, 2123 (2012).
[26] R. Baer and D. Neuhauser, J. Chem. Phys. **137**, 051103 (2012).
[27] D. Neuhauser, E. Rabani, and R. Baer, J. Chem. Theor. Comp **9**, 24 (2012).
[28] D. Neuhauser, E. Rabani, and R. Baer, J. Phys. Chem. Lett. **4**, 1172 (2013).
[29] R. Kosloff, J. Phys. Chem. **92**, 2087 (1988).
[30] S. Goedecker and L. Colombo, Phys. Rev. Lett. **73**, 122 (1994).
[31] Y. H. Huang, D. J. Kouri, and D. K. Hoffman, Chem. Phys. Lett. **243**, 367 (1995).
[32] R. Baer and M. Head-Gordon, J. Chem. Phys. **107**, 10003 (1997).
[33] L. W. Wang, Phys. Rev. B **49**, 10154 (1994).
[34] J. P. Perdew and Y. Wang, Phys. Rev. B **45**, 13244 (1992).
[35] N. Troullier and J. L. Martins, Phys. Rev. B **43**, 1993 (1991).
[36] P. Pulay, Chem. Phys. Lett. **73**, 393 (1980).